\documentstyle[preprint,aps]{revtex}
\begin{document}
\draft

\newcommand{\dd}{^{\circ}}

\title{Polymeric alkali fullerides are stable in air}

\author{Daniel Koller, Michael C. Martin, Peter W. Stephens, Laszlo Mihaly}
\address{Department of Physics, State University of New York at Stony Brook,
NY 11794-3800, USA.}
\author{Sandor Pekker}
\address{Research Institute for Solid State Physics, POB 49,
H-1525 Budapest, Hungary}
\author{Andras Janossy}
\address{Institute of Physics, Technical University of Budapest,
H-1521, Budapest, Hungary}
\author{Olivier Chauvet and Laszlo Forro}
\address{Laboratoire de Physique des Solides Semicrystallines, IGA,
Department of Physics, EPFL, 1015 Lausanne, Switzerland}

\date{16 May 1994; to be published Appl. Phys. Lett. 20 Feb 1995}
\maketitle

\begin{abstract}
     Infrared transmission, electron spin resonance, and X-ray diffraction
     measurements show unambiguously that RbC$_{60}$ and KC$_{60}$ are stable
     in air, in contrast to Rb$_{6}$C$_{60}$ which decomposes rapidly upon
     exposure.  The specimens studied transform into pure C$_{60}$ and other
     byproducts when heated above $100\dd $C, approximately the temperature
     of the orthorhombic-fcc phase transition.  The stability of these
     compounds raises the possibility of applying them as protective
     layers for the superconducting fullerides.
\end{abstract}

     The compounds of alkali metals (A) with fullerenes (C$_{60}$) are the
subject of considerable recent interest.  Superconductivity was discovered
in A$_{3}$C$_{60}$\cite{1,2}, and a polymeric chain structure\cite{3,4}
with metallic properties\cite{5,6} was seen in A$_{1}$C$_{60}$.  Other
fullerides of composition A$_{4}$C$_{60}$ and A$_{6}$C$_{60}$ have also been
studied\cite{2}.  Early investigations established that some of these
materials rapidly decompose if exposed to air, and extreme care has been
taken to treat all samples in an inert atmosphere\cite{2}.  Yet here we
report Infrared (IR) spectroscopy, Electron Spin Resonance (ESR) and X-ray
diffraction experiments, indicating that two fullerides, RbC$_{60}$
and KC$_{60}$ are stable in air.

     The A$_{1}$C$_{60}$ compounds have an fcc (rocksalt) structure at high
temperature\cite{7}, and upon slow cooling they undergo a reversible first
order phase transition to an orthorhombic state, where the C$_{60}$ molecules
are chemically bonded to form linear chains\cite{3,4}.  This ``polymeric" state
is thermodynamically stable at room temperature for RbC$_{60}$.  Phase
separation of KC$_{60}$ to K$_{3}$C$_{60}$ and pure C$_{60}$ was reported
by several authors\cite{7,8,9}.  However, for the slowly cooled samples
in this study we consistently found large amounts of polymeric phase of both
RbC$_{60}$ and KC$_{60}$ by ESR, X-ray diffraction\cite{4} and IR
spectroscopy.  Further investigations are clearly needed in this respect, but
the basic conclusions of the present work are not expected to change.

     In the present study three separately prepared sets of samples were
investigated.  RbC$_{60}$ thin films (specimen S1) were made on Si substrates
under vacuum in a special sample cell\cite{10}.  Doping was performed by
exposing the films to Rb vapor.  During doping the temperature of the
film was $225\dd $C and the composition was monitored {\it in-situ} by
recording IR spectra, and following the evolution of the F$_{1u}(4)$ molecular
vibration\cite{9,11}.  Polycrystalline KC$_{60}$ (specimen S2) was made by
co-evaporation of stoichiometric amounts of the constituent materials in a
sealed tube, placed in a gradient furnace\cite{12}.  Crystals of typical
size $\sim$0.1mm grew at a temperature of about $300\dd $C, while the C$_{60}$
and K were kept at $600\dd $C and $150\dd $C, respectively.  During the course
of the experiments samples S1 and S2 were first investigated under vacuum or
inert atmosphere, and were later exposed to air.  A part of the pristine
S2 sample was further processed to produce iodine treated KC$_{60}$.  The
iodination was carried out in a glove box by immersing the co-evaporated
crystals to a dilute solution of iodine in toluene for four days.  The
mixture was then repeatedly rinsed with toluene, ethanol and pentane in air.
Specimen S3 was the insoluble reside, left behind after this process.  This
sample was stored in air for about a month before the measurements reported
here were completed.

     IR transmission measurements were carried out on samples S1 and S3.
First the spectrum of the pristine sample S1 was recorded.  The low overall
transmission and the characteristic resonance structure\cite{6,13} of
RbC$_{60}$ were clearly visible (lower curve of Fig. 1).  The IR spectrum
indicates that S1 also contains measurable amounts of Rb$_{6}$C$_{60}$ and
pure C$_{60}$.  Then the hermetically sealed sample cell was opened to air.
The spectrum taken immediately afterwards ($\sim$10min) shows the same
RbC$_{60}$ resonance lines, slightly increased transmission, and the absence
of the Rb$_{6}$C$_{60}$ impurity phase (Fig. 1, upper curve).  Therefore we
conclude that Rb$_{6}$C$_{60}$ was destroyed and RbC$_{60}$ is stable.  The
small increase in overall transmission is most likely due to the removal of a
small amount of metallic Rb$_{3}$C$_{60}$ from the pristine sample.  The
corresponding Rb$_{3}$C$_{60}$ resonance line, positioned at 1364cm, is
expected to be below the noise level of the experiment.  Pure C$_{60}$, most
likely produced in the chemical reactions after air exposure, has a much weaker
IR signal than Rb$_{6}$C$_{60}$ or Rb$_{3}$C$_{60}$\cite{9,11,14} and does
not change the magnitude of the existing C$_{60}$ signal by any appreciable
amount.

     The IR spectrum of the iodine treated sample S3 is also dominated by the
polymeric KC$_{60}$ signal (Fig. 2).  This specimen was used to demonstrate
that the compound is unstable if moderately heated.  Upon heating the
transmission increased dramatically, the split F$_{1u}(4)$ line of
RbC$_{60}$ at 1387cm$^{-1}$ and 1406cm$^{-1}$ disappeared and a strong
C$_{60}$ line appeared; i.e. the sample transformed into a mixture consisting
of pure C$_{60}$ and other compounds with no strong IR resonances in the
measured frequency range.  The decomposition starts
very slowly below $100\dd $C, but it is complete at T=$200\dd $C. The sample
was heated in several steps in the IR spectrometer, and it reached $200\dd $C
in about three hours.  As illustrated by the uppermost
spectrum at room temperature, the decomposition is irreversible.

     The co-evaporated crystals (S2), as well as S3, were used in the ESR
measurements. KC$_{60}$ and RbC$_{60}$ have a characteristically narrow ESR
line\cite{5,15}. The ESR spectrum of sample S1 was recorded at room
temperature, with the sample sealed in a quartz tube.  The tube was then
opened to air, and the ESR signal was measured for several days.  Figure 3
shows that the signal intensity a.) did not change after opening the sample
to air and b.) was independent of time within the accuracy of the measurement.
The width of the signal was also unchanged.  We emphasize that the most
likely by-products of the disintegration of KC$_{60}$ (C$_{60}$ and K$_{2}$O
or KOH) have no ESR signal.  If free radicals or other paramagnetic materials
are produced, then it is extremely unlikely that their signals mimic the
KC$_{60}$ signal so accurately. The ESR signal of sample S3 is also
characteristic of polymeric KC$_{60}$, and the magnetic susceptibility
remains constant down to 50K, indicating the metallic nature of electrons,
similar to pristine polymeric KC$_{60}$\cite{5}.

     X-ray spectroscopy was also performed on the iodine treated sample, S3.
The diffraction spectrum (Fig. 4.) shows an enhanced background, possibly due
to a non-crystalline component, but the majority of the crystalline part is
clearly orthorhombic KC$_{60}$\cite{4}.  All three of the probes, X-ray,
ESR and IR spectroscopy, sensitive to various physical properties of the
sample, show that the air-exposed specimens are mostly polymeric
A$_{1}$C$_{60}$.

     The experimental data presently available to us is not sufficient to
completely explain the unexpected stability of polymeric
alkali metal fullerides.  Some C$_{60}$ salts with a C$^{-}$ anion, like the
(TDAE)C$_{60}$ or the chromium(III) porfirin salt, were found to be sensitive
to air\cite{16,17}, while others, like (tetraphenyl phosphonium)C$_{60}$,
are stable\cite{17}.  In principle, stability is observed if the material
is in a thermodynamically favored configuration (like gold in air) or it
may be covered by a protecting layer (like aluminum in air).  The linear
chain structure\cite{4} and morphology\cite{12} of the AC$_{60}$ compounds
suggest a more exotic possibility:  the long C$_{60}$ polymer chains inhibit
the diffusion of the alkali metal or the oxygen, effectively leading to a much
slower one-dimensional diffusion and hence a greatly enhanced lifetime.

     The existence of a stable alkali metal fulleride may have far-reaching
consequences for the study and application of fullerides.  Sample preparation
procedures and measurements are much simpler if the specimens can be freely
moved and exchanged in air.  Investigation of the DC electrical properties
will clarify if inter-grain potential barriers are present, possibly inhibiting
the electronic transport.  In any case, thermal, magnetic and optical studies
are greatly simplified.  The stable RbC$_{60}$ or KC$_{60}$ polymer may
have applications as a protective skin for the
superconducting K$_{3}$C$_{60}$ or Rb$_{3}$C$_{60}$ material. In fact, the
present results may shed some light on early reports of unexpected persistence
of the superconducting Meissner signal in some air exposed A$_{3}$C$_{60}$
samples\cite{18}.  We hypothesize that a fraction of the material in those
samples was doped incompletely, leading to the formation of A$_{1}$C$_{60}$,
which then encapsulated the A$_{3}$C$_{60}$.

\acknowledgements
   This work has been supported by the NSF grant DMR 9202528
Swiss National Foundation for Scientific Research grants
No. 2100-037318 and 7UNPJ038426 US Hungarian Joint Fund. JF225
and OTKA grants 2932, 2979 and T422.

\narrowtext
\begin{figure}
\caption{
IR transmission spectra of a thin Rb-doped C$_{60}$ film (specimen S1).  Inset
shows the lower two vibrational modes on an expanded scale.  The sample was
cooled at a rate of $\sim$$10\dd $C/min to room temperature from the
preparation temperature.  The first spectrum (lower curve) was measured on the
pristine film in vacuum.  The lines at 1387cm$^{-1}$ and 1406cm$^{-1}$ derive
from the F$_{1u}(4)$ molecular vibration, split by the polymerization.  The
F$_{1u}$(2) and F$_{1u}$(4) derived resonances corresponding the
Rb$_{6}$C$_{60}$ are also clearly visible.  When the sample is exposed to air
(upper curve), the Rb$_{6}$C$_{60}$ is destroyed, while the RbC$_{60}$
remains.}
\label{fig1}
\end{figure}

\begin{figure}
\caption{
IR transmission of the iodine treated KC$_{60}$ powder (specimen S3) in a KBr
pellet.  The spectral features before heating indicate that the majority of
the sample is KC$_{60}$.  Upon heating, the sample decomposes.  The right side
scale, which belongs to the lower two curves, is expanded by a factor of ten
relative to the left side scale.}
\label{fig2}
\end{figure}

\begin{figure}
\caption{
Time dependence of the ESR intensity of a polycrystalline KC$_{60}$ sample
(specimen S2) after exposure to air.}
\label{fig3}
\end{figure}

\begin{figure}
\caption{
X-ray diffractogram of the iodine exposed KC$_{60}$ (specimen S3), as compared
to a pristine polycrystalline sample, used in the detailed Rietveld analysis of
the polymeric structure (Ref. 4).}
\label{fig4}
\end{figure}

\end{document}